\documentclass[%
 aip,
 amsmath,amssymb,
 reprint,%
]{revtex4-1}

\usepackage{graphicx}
\usepackage{dcolumn}
\usepackage{bm}

\usepackage[utf8]{inputenc}
\usepackage[T1]{fontenc}
\usepackage{mathptmx}
\usepackage{etoolbox}
\usepackage{siunitx} 

\makeatletter
\def\@email#1#2{%
 \endgroup
 \patchcmd{\titleblock@produce}
  {\frontmatter@RRAPformat}
  {\frontmatter@RRAPformat{\produce@RRAP{*#1\href{mailto:#2}{#2}}}\frontmatter@RRAPformat}
  {}{}
}%
\makeatother
\begin{document}

\preprint{AIP/123-QED}

\title{Characterization of foam-filled hohlraums for inertial fusion experiments}

\author{S. Iaquinta}
\email{sam.iaquinta@physics.ox.ac.uk}
 \affiliation{Department of Physics, University of Oxford, Parks Road, Oxford OX1 3PU, United Kingdom.}
 
\author{P. Amendt}
\author{J. Milovich}
\author{E. Dewald}
\author{L. Divol}
\author{O. Jones\textsuperscript{\textdagger}}
\author{L. Suter}
\author{R. Wallace}

\affiliation{Lawrence Livermore National Laboratory, Livermore, California 94550, USA.}%

\author{R. Bingham}
\affiliation{$^3$STFC Rutherford Appleton Laboratory, Chilton, Didcot OX11 OQX, United Kingdom.}
\affiliation{Department of Physics, SUPA, University of Strathclyde, Glasgow G4 0NG, United Kingdom.}

\author{S. Glenzer}
\affiliation{SLAC National Accelerator Laboratory, 2575 Sand Hill Road, Menlo Park, California 94025, USA.}

\author{G. Gregori}
\affiliation{%
Department of Physics, University of Oxford, Parks Road, Oxford OX1 3PU, United Kingdom.
}%

\date{\today}

\begin{abstract}
On the path towards high-gain inertial confinement fusion ignition, foams are being considered to tamp the hohlraum wall-motion, and mitigate laser backscattering from Stimulated Raman Scattering (SRS) and Stimulated Brillouin Scattering (SBS). Here we present the results from an experimental campaign on foam-filled hohlraums conducted at the OMEGA laser facility. SiO$_2$ foam-fills, with densities as low as 1 mg/cm$^3$, successfully reduce the gold wall expansion, with laser backscattering comparable to gas-fills.
\end{abstract}

\maketitle

Recent fusion results from the National Ignition Facility (NIF) \cite{NIF_gain_1.5, NIF_gain_0.7} have increased interest in inertial fusion energy. The NIF results were obtained using the indirect drive approach to Inertial Confinement Fusion (ICF), where a cryogenic deuterium-tritium (DT) fuel capsule is placed inside a high-Z (typically gold) cylinder called a hohlraum. High intensity laser beams are focused onto the hohlraum walls, converting the laser light into x-ray radiation. The x rays then heat the low Z outer layer of the capsule and compress it through ablation pressure to create a high temperature hotspot, triggering nuclear fusion reactions. In order to reach the thermonuclear conditions required to ignite the fuel, the capsule must converge by a factor of $\sim$ 30. Hence, small defects and asymmetries can have a significant impact on the quality of the implosion, and can be amplified by instabilities during compression such as Rayleigh-Taylor \cite{Rayleigh-Taylor} and Richtmyer–Meshkov \cite{Richtmyer_Meshkov} instabilities. Another approach to laser fusion is to focus the lasers directly onto the capsule known as direct drive \cite{Direct_drive_campbell}. As opposed to direct drive, the indirect drive approach allows for a smoother and more symmetrical drive; however, several challenges still persist and impose limitations on the path to high-gain ignition. Namely, the expansion of the laser-heated hohlraum walls results in regions of overdense plasma preventing some of the laser beams reaching the inner region. This reduces the conversion of laser energy into x-rays and relocates the laser hot spots, resulting in x-ray drive asymmetries. In large laser facilities such as the National Ignition Facility (NIF) \cite{NIF_Glenzer_2004}, laser beams are divided into outer and inner beams for symmetry purposes. A specific limitation imposed by this expansion of the gold wall is the growth of a gold \textit{bubble} at the outer-beam laser deposition locations that blocks the inner beams from reaching the hohlraum equator at late time. Hohlraums are therefore typically filled with gas to tamp the wall motion and reduce asymmetries in the radiation drive. However, high density gas fills ($\gtrsim $ 0.85 mg/cm$^3$) result in significant amounts of laser backscatter \cite{Jones_2016} due to Laser Plasma Instabilities (LPI) that develop in the ionized gas fill. On the other hand, lowering the gas-fill density reduces symmetry control. Foams are being considered as an alternative solution to reduce the wall expansion with minimal to no increase in LPI. In this Letter we present the first experimental results on the performance of SiO$_2$ foam-filled hohlraums at the Omega Laser Facility \cite{OMEGA}.

     \begin{figure}[b]
                \centering
                \begin{tabular}{cc}
                (a) \\
                \includegraphics[width=0.40\textwidth, trim = {0 0.5cm 0 1cm}, clip]{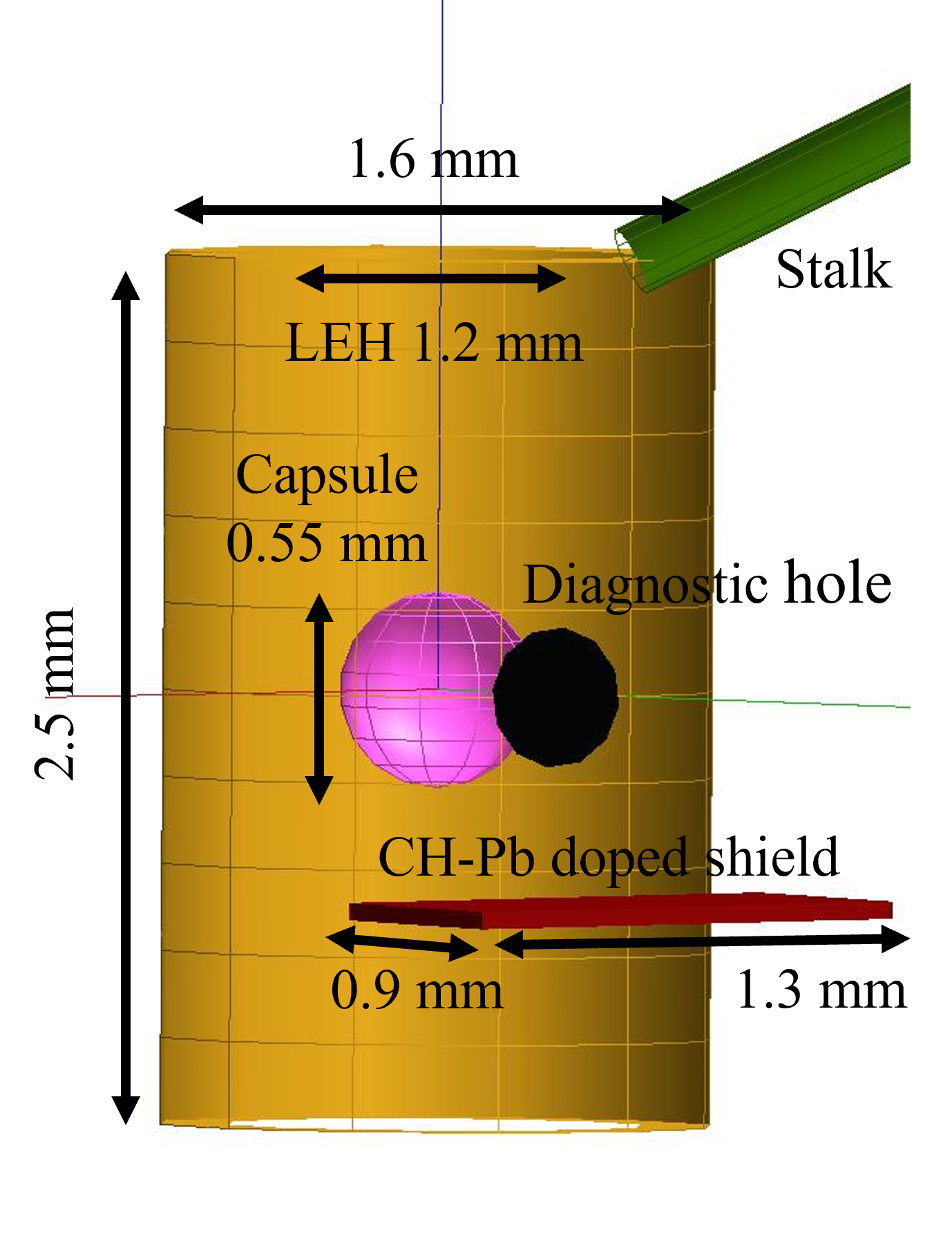}\\
                (b)\\
                \raisebox{0.1cm}{\includegraphics[ width=0.40\textwidth, trim = {0.7cm 0.3cm 0.9cm 0}, clip]{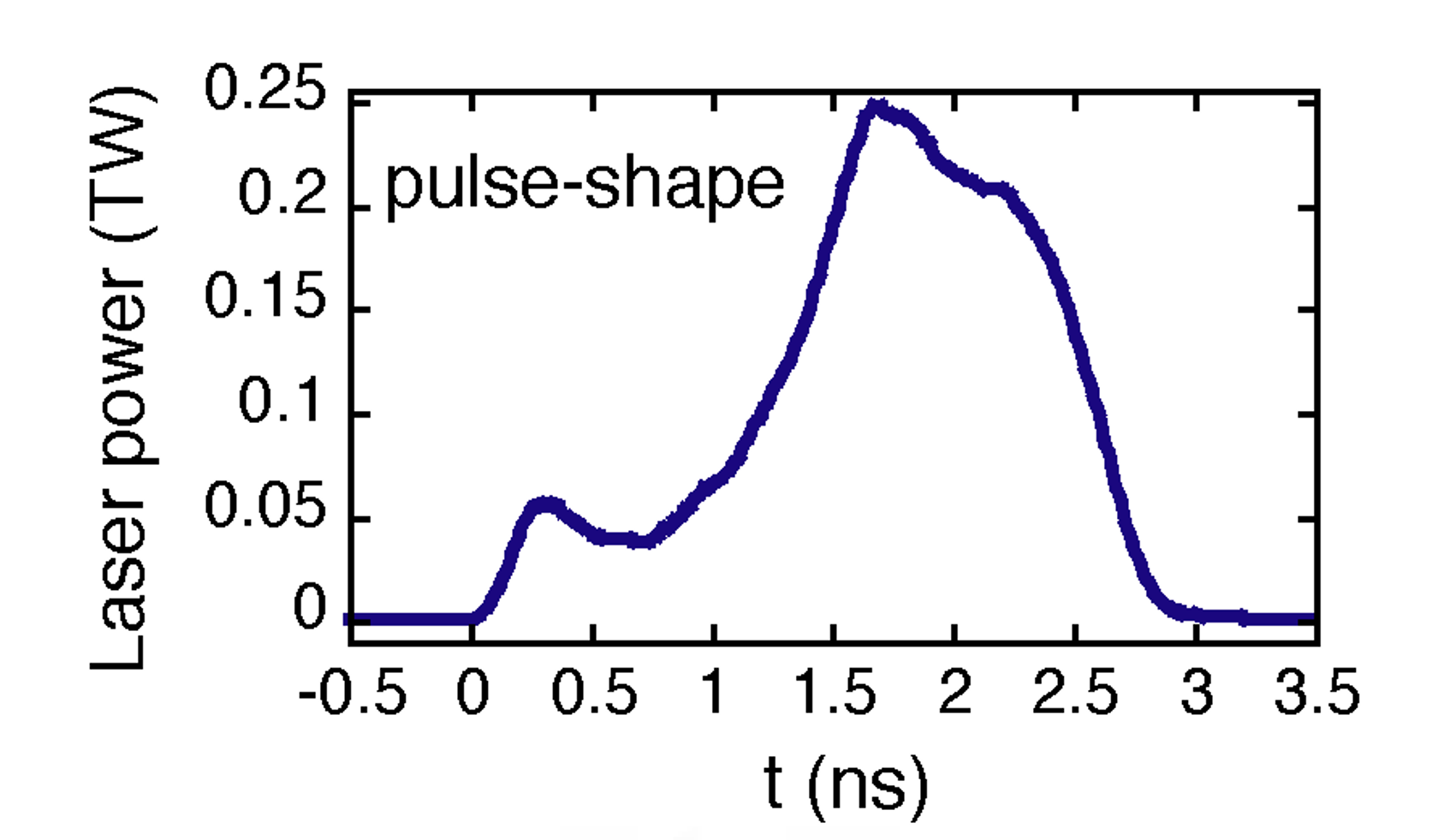}}
                \end{tabular}
                \caption{(a) Design and dimensions of the target used in almost all shots. The diagnostic hole was patched with 25 $\mu$m Be and 2 $\mu$m Ta. (b) Shape of the laser pulse. It is a $\sim$3 ns pulse, consisting of a 1 ns foot followed by a rising edge with a maximum at around 1.6 ns, and by an approximately 1 ns flatop and falling edge. The average total peak power was around 10 TW and the intensity at the LEH was approximately $\SI{9e14}{W\per\cm\squared}$.}
                \label{visrad_omega_target}
    \end{figure}

The ionisation of the foam or gas fill, as well as the ablation of the capsule and gold walls, produce a flowing plasma. The interaction of this plasma with the incoming high-intensity laser beams may lead to the generation of several laser plasma instabilities \cite{LPI_NIF_2010, Kruer}. Having a good understanding of LPI for the proposed hohlraums is therefore crucial to optimise ICF experiments. In both direct and indirect approaches to ICF, the main LPIs of interest are Stimulated Raman Scattering (SRS) and Stimulated Brillouin Scattering (SBS) \cite{scattered_processes_Forslund_1975, LPI_relevant_to_ICF, LPI_relevant_to_ICF2, LPI_relevant_to_ICF3, LPI_relevant_to_ICF4}. 
These three-wave parametric instabilities can only occur provided that the Manley-Rowe relations \cite{Manley-Rowe} are respected

\begin{equation}
\label{wave matching}
    \displaystyle \begin{split}\omega_{0} = \omega_s + \omega_2\\
    \boldsymbol{k_0} = \boldsymbol{k_s} + \boldsymbol{k_2}.
    \end{split}
\end{equation}

\noindent where $\omega_{0, s}, \boldsymbol{k_{0, s}}$ are the frequency and wavenumber of the incident laser wave and scattered wave respectively, and $\omega_{2}, \boldsymbol{k_{2}}$ are the frequency and wave number of a density perturbation such as one associated with an electron plasma wave (SRS) or with an ion acoustic wave (SBS).

SRS is of particular concern for ICF as the resulting enhanced electron plasma wave can produce hot electrons via Landau Damping \cite{Landau_damping}. Hot electrons are detrimental to ICF as they can preheat the capsule fuel \cite{preheat,preheat2}, reducing its convergence and consequently the thermonuclear yield. SBS is another concern for ICF as the backscatter can go back up the laser chain and damage the optics. Furthermore, both SRS and SBS can lead to asymmetries in the x-ray drive if the backscatter levels are sufficiently high. For laser intensities $I_0 > \SI{e14}{W\per\cm\squared}$, temperatures of a few keV and densities around $n_e = 0.1 n_c$, where $n_c$ is the critical density, the SBS growth time $\gamma^{-1}$ is larger than that of SRS by a factor of around 4, but both have $\gamma^{-1} \lesssim 1$ ps. Hence, SRS and SBS have enough time to develop in the nanosecond timescale of ICF experiments.
 
Foams are currently being considered \cite{foam_imaging, foam_homogenization,foam_oggie_aps,foam_ion_heating,foam_Gus_kov_1997,foam_oggie} instead of gas fills, as a complementary approach, to tamp the wall-motion whilst also mitigating LPIs. Open-celled foams are solid structures separated by pockets of gas and typically have an inhomogeneous density structure. Currently it is possible to manufacture foams with densities as low as 1 mg/cc\cite{foam_oggie}, which is ideal for an ICF target since such a density would correspond to only about $3.3\%$ of the critical density for a \mbox{$351$ nm} pulse. Furthermore, experimental measurements of laser backscattering from foams have been lower than predicted due to preferential ion heating \cite{foam_Gus_kov_1997,foam_ion_heating,foam_oggie_aps, foam_ion_electron_heating, subgrid}. Highly resolved radiation-hydrodynamic computer simulations of the foam structure were able to reproduce the reduction of SBS due to this mechanism \cite{foam_Milovich_2021}. Foams also offer the possibility of introducing dopants \cite{foam_dopant_SRS, foam_dopant_SBS}, even at cryogenic temperatures, which have been shown to further suppress backscattering \cite{foam_dopant_SRS, foam_dopant_SBS}. Although a number of experiments on foam liners have been performed \cite{foam_liner, foam_liner_laser_imprint,foam_liner_laser_imprint2,foam_liner_Moore}, we only consider in this Letter solid liners and hohlraums entirely filled with foam. 

      \begin{figure}[t]
                \centering
                \begin{tabular}{cc}
                (a) \\
                \includegraphics[width = 0.49\textwidth, trim = {0cm 0cm 0cm 0.2cm}, clip]{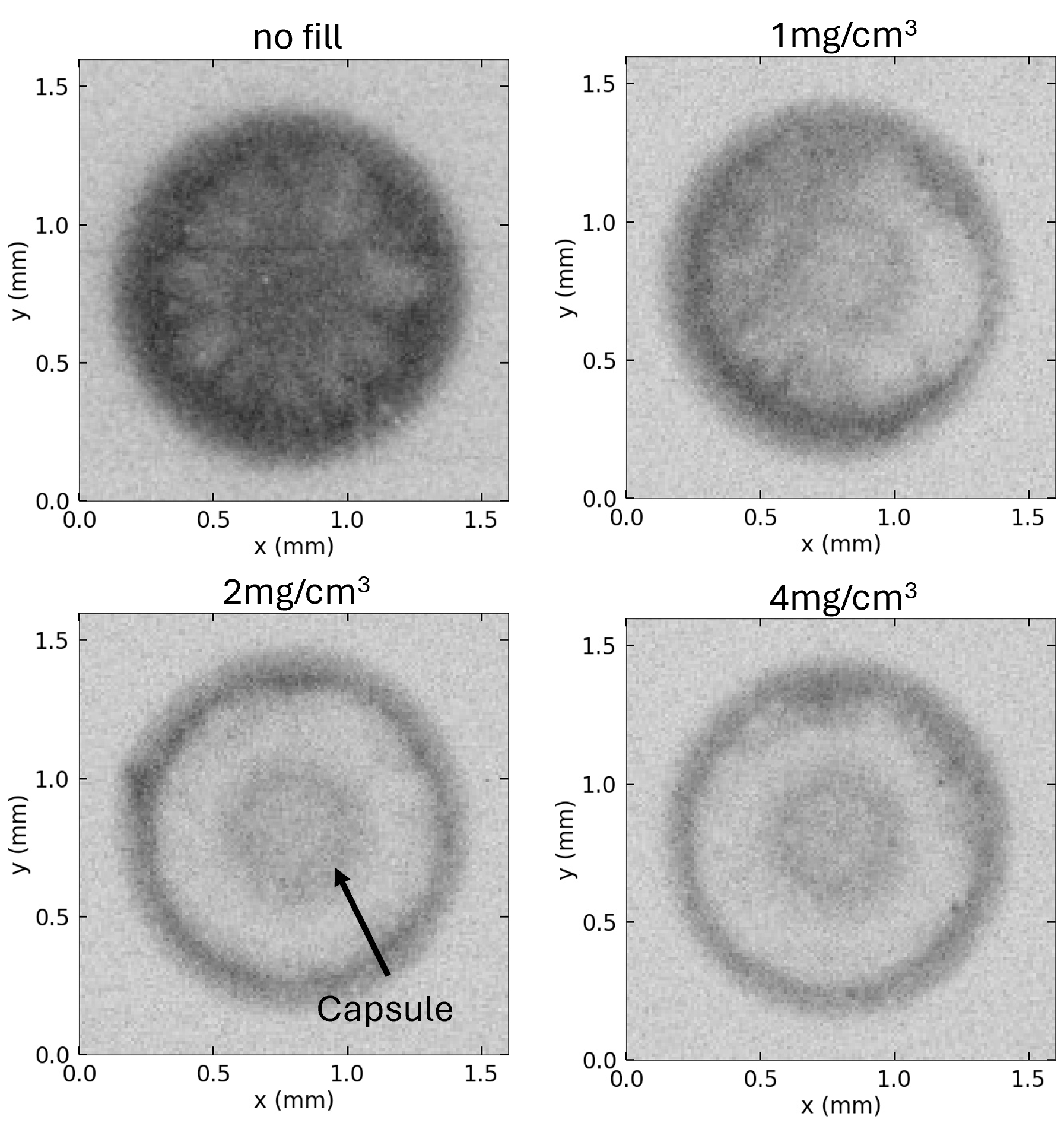} \\
                (b) \\
                \includegraphics[width = 0.45\textwidth]{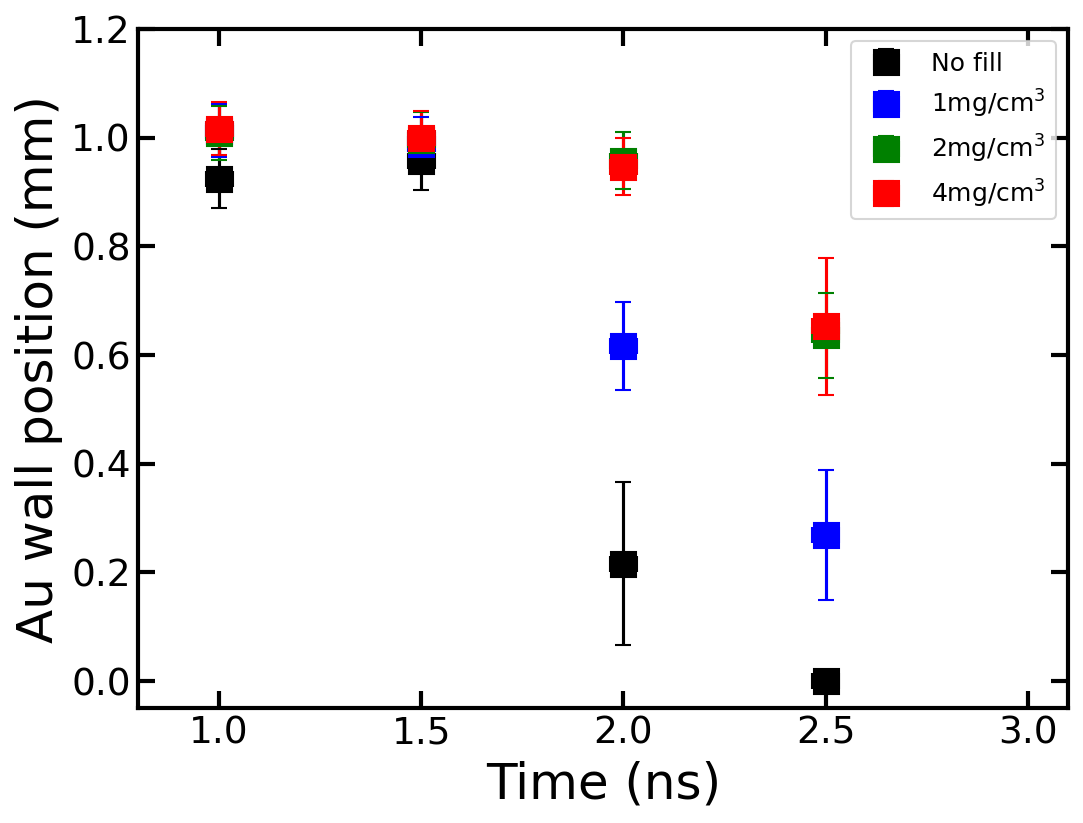}
                \end{tabular}
                \caption{(a) Soft x-ray images of the target through LEH at 2 ns for different foam densities. The empty hohlraum appears almost completely filled with Au blow-off plasma whereas the other images show successful wall-tampering from the foam. (b) The evolution over time of the wall-motion as the gold \textit{bubble} expands for different densities. Closure can be seen from around 2.5 ns onward for the empty hohlraum whereas foams show reduced wall expansion due to the back-pressure provided, with high densities yielding slower expansions.} 
                \label{LEH}
    \end{figure}

      \begin{figure*}[t]
                \centering
                \begin{tabular}{cc}
                (a) & (b) \\
                \includegraphics[width = 0.58\textwidth, trim = {0cm 0 0cm 0}, clip]{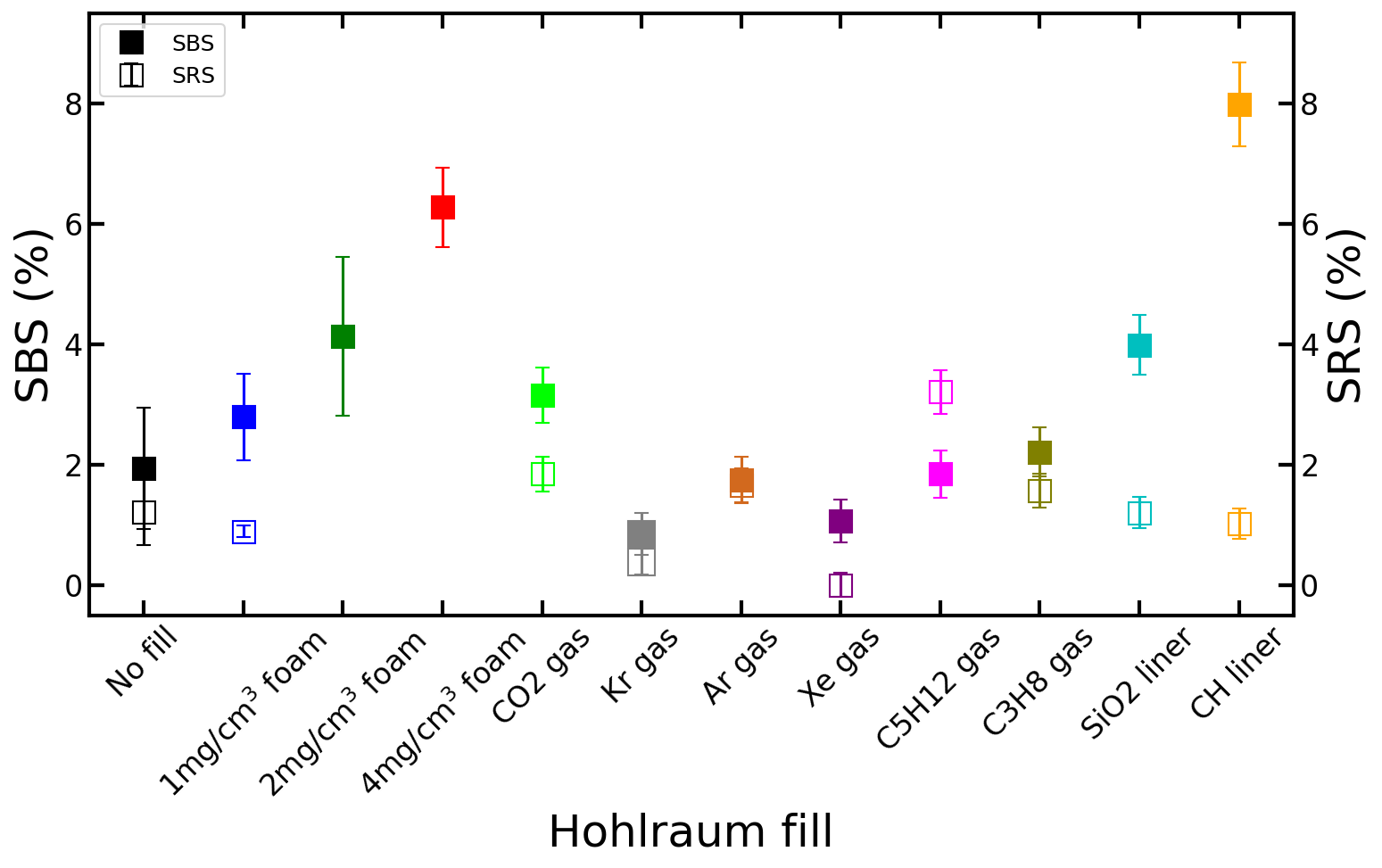} & \raisebox{1cm}{\includegraphics[width = 0.40\textwidth]{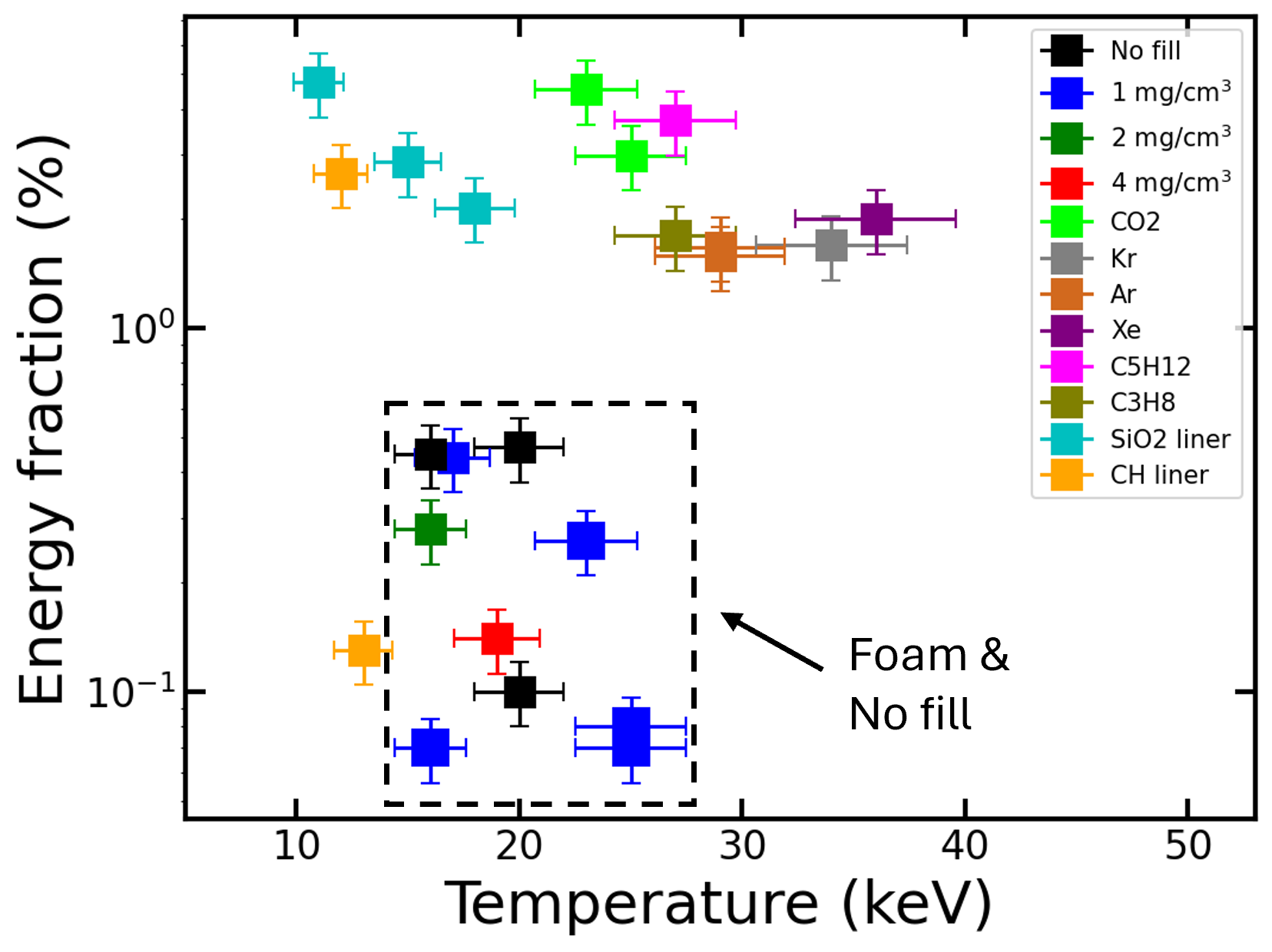}} \\
                (c) & (d) \\
                \includegraphics[width = 0.61\textwidth, trim = {0cm 0cm 0cm 0.0cm}, clip]{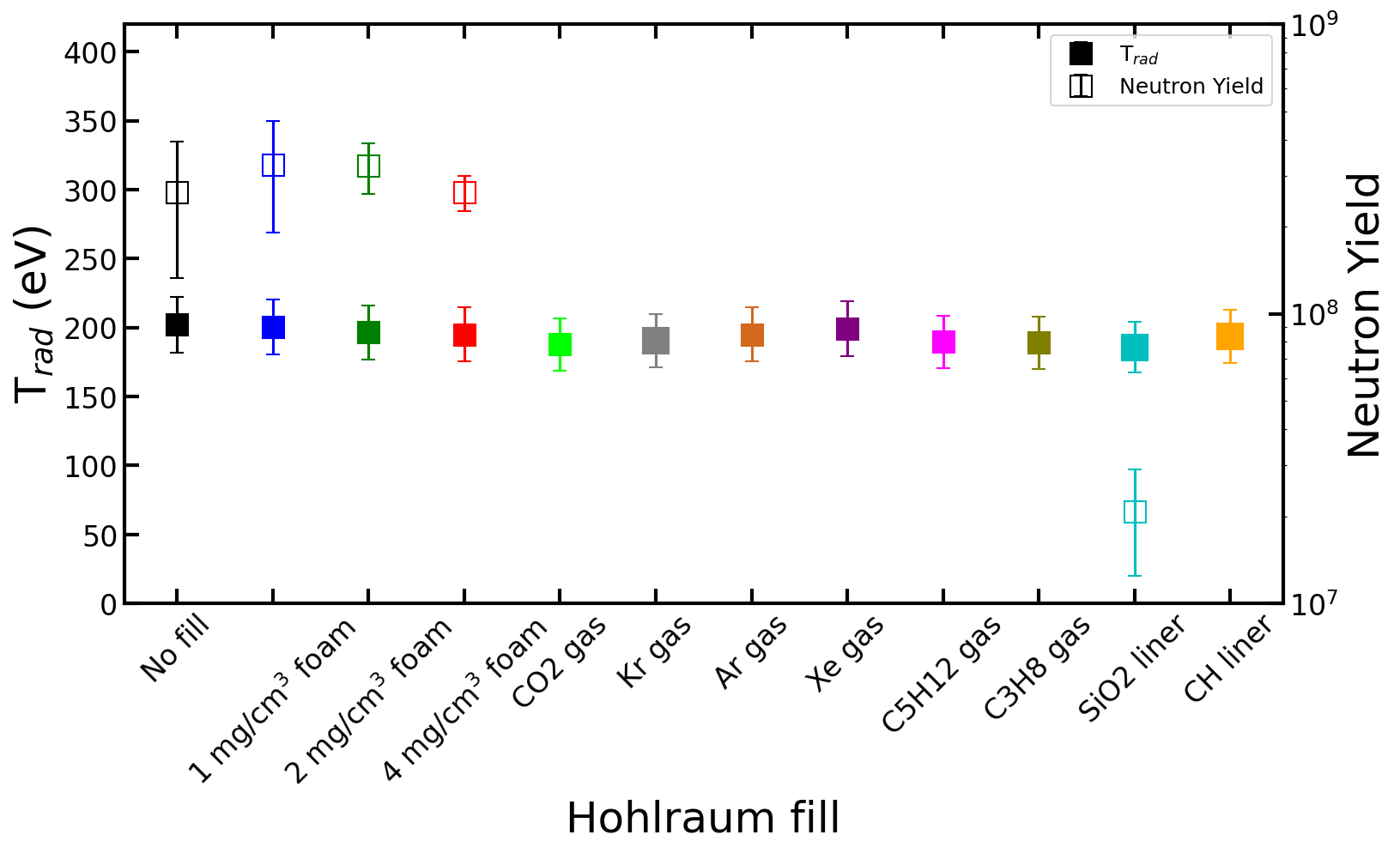} & \includegraphics[width = 0.35\textwidth, trim = {0cm 0cm 0cm 0cm}, clip]{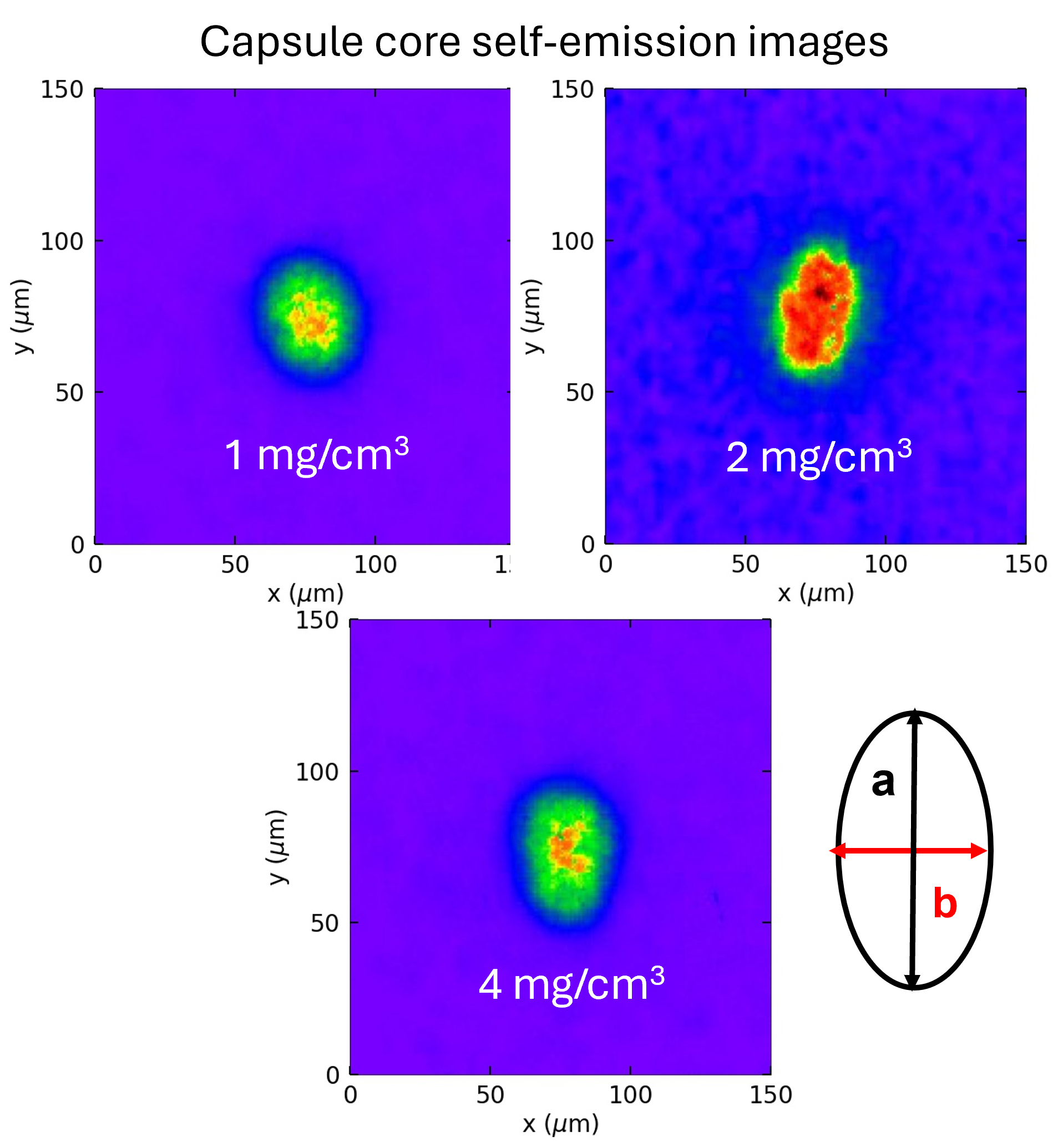}

                \end{tabular}
                \caption{(a) Stimulated Brillouin Scattering (SBS) and Stimulated Raman Scattering (SRS) measurements from FABS show similar amounts of backscattering for low density (1 mg/cm$^3$) foam-filled, gas-filled and empty hohlraums. Higher density foams and liners show increased levels of SBS. (b) The fraction of the total laser energy that is converted into the generation of hot electrons is plotted against the average temperature of those electrons. Foams show lower energy conversion than gases and liners, with similar average temperatures. (c) Measurements from Dante show similar performance in radiation temperature for all fills. Similarly, neutron yield measurements show comparable implosion performance to empty hohlraums. Liners however perform significantly worse with regards to their neutron yield. (d) X-ray images of the capsules' self-emission for different foam densities. The implosion symmetry was quantified by the capsule's aspect ratio $a/b$, where $a$ is the vertical axis and $b$ the horizontal axis. Higher foam densities have a more prolate symmetry, which, given that the hohlraum axis is horizontal, suggests potential cross beam energy transfer toward the outer cones.} 
                \label{fig3}
    \end{figure*}

Here we present the results of an experimental campaign on foam-filled hohlraums conducted at OMEGA in the Laboratory for Laser Energetics (LLE). The main objective of the experiments is to assess the relative advantages and disadvantages of SiO$_2$ foams when compared to other hohlraum fills (i.e., several gas-fills, CH and SiO$_2$ solid liners, and no-fill hohlraums). Three foam densities were tested, namely 1 mg/cm$^3$, 2 mg/cm$^3$ and 4 mg/cm$^3$. The gas fill pressure for all gases was between 0.9 atm and 1 atm, the CH liners were 0.5 $\mu$m thick and the SiO$_2$ liners were 0.25 $\mu$m thick. The target design was almost identical for every shot and is illustrated in Fig. \ref{visrad_omega_target}(a). The Au hohlraum was 2.5 mm long, \mbox{with a 1.6 mm diameter} and a 1.2 mm Laser Entrance Hole (LEH) diameter. The capsule used in these experiments was a 220 $\mu$m inner radius shell filled with 10 atm of deuterium and 0.025 atm of argon dopant. The CH shell was 30 $\mu$m thick and doped with 1\% atomic fraction of germanium.  
A total of 40 beams, spread into 3 cones, with a combined energy ranging from 12 kJ to 14 kJ were focused onto the target. The 5:1 foot-to-main ratio pulse shape PS26 \cite{ps26} was used for all shots and is shown in Fig. \ref{visrad_omega_target}(b). The beam quality was improved via smoothing by spectral dispersion \cite{SSD} and distributed polarization rotators \cite{DPR_rochester}. 
The primary diagnostics suite included a soft x-ray framing camera to image the wall-motion through the LEH, another x-ray framing camera to capture the argon emission from the capsule, two Full Aperture Backscatter Stations (FABS) \cite{FABS} to measure the amount of laser backscattering, the time-resolved, absolutely calibrated spectrometer Dante \cite{DANTE} to measure the x-ray drive radiation temperature, a hard x-ray detector \cite{hard_x-ray2} to measure the production of hot electrons, and  scintillators to measure the neutron yield. 



Several soft x-ray (760 eV) images of the LEH were taken at different times, between 1 ns and 3 ns, with a framing camera to monitor the hohlraum wall motion as shown in Fig. \ref{LEH}. It is clear that the Au blow-off is delayed in foam filled targets compared to the empty one. The evolution over time of the wall expansion was measured from these images. Empty hohlraums became completely filled with overdense Au plasma from $\sim$2.5 ns onward, blocking the laser beams from propagating. Foams however, show clear evidence of wall tamping, delaying the wall motion by approximately 0.5 ns for the 1 mg/cm$^3$ case and even longer for the 2 mg/cm$^3$ and 4mg/cm$^3$ foams.


The amount of laser backscattering via SBS and SRS was measured using FABS. The uncollected backscattered light that did not enter the FABS collection lens was not accounted for in the data. Since small amount of light was seen on the near-bacskcatter imaging screens put around the collection optics, the data shown here must be understood as a lower bound only. As indicated in Fig. \ref{fig3}(a), low density foams (i.e., 1 mg/cm$^3$) exhibit similar amounts of SBS as empty hohlraums and most gas fills, with slightly less SRS than all gases except Kr and Xe. However, the amount of SBS (in \% of incoming laser energy) increases with foam density by approximately 1.15\% per mg/cm$^3$ according to Fig. \ref{fig3}(a). Liners have also higher SBS with CH being the highest.       


The temperature of the hot electrons generated via SRS, as well as their corresponding total laser energy fraction, was measured and is shown in Fig. \ref{fig3}(b). Only low energy channels of the hard x-ray detector were used, hence \mbox{high energy ($> 100$ keV)} hot electrons were not considered. Foams and empty holraums \mbox{show lower} laser energy conversion into hot electrons, as well as slightly lower temperature ($\approx 20$ keV) compared to gases. Most liners exhibit similar hot electron temperatures to foams but higher total energy.


Radiation temperature measurements of the x-ray drive were taken with Dante and are plotted in Fig. \ref{fig3}(c). Similar radiation temperatures were observed for all fills, around 200 eV, indicating that foams do not reduce laser energy conversion into radiation drive. 

Similarly, the nuclear performance of the capsule implosion for hohlraums using foam fills was comparable to that of the empty hohlraums, between $10^8$ and $10^9$ neutrons produced, as shown in Fig. \ref{fig3}(c), indicating good overall implosion performances for foam-filled hohlraums. On the other hand, targets using hohlraums with SiO$_2$ liners were found to perform significantly poorly, with neutron yields $Y < 10^8$.

Finally, the implosion symmetry was measured from x-ray images of the argon self-emission from the capsule using an x-ray framing camera with a 0.127 mm beryllium filter. Fig. \ref{fig3}(d) shows said images for the 1 mg/cm$^3$, 2 mg/cm$^3$ and 4 mg/cm$^3$ foam-filled targets. The aspect ratios $a/b$ of the capsules' core were measured from these self-emission images, where $a$ is the vertical axis of the capsule and $b$ the horizontal axis. It is important to note that the hohlraum axis is horizontal and that the symmetry and aspect ratios are measured with respect to it. The best implosion symmetry was obtained with the low density 1 mg/cm$^3$ foam target ($a/b = 1.1 \pm 0.1$). This round symmetry could be a result of the delay of the gold wall blow-off by the foam, allowing for the inner beams to reach the equator of the hohlraum, yielding a more symmetrical drive. However, increasing the foam density seems to affect the symmetry with aspect ratios of $1.54 \pm 0.15$ and $1.39 \pm 0.14$ for the 2 mg/cm$^3$ and 4 mg/cm$^3$ targets respectively. This more prolate symmetry (with respect to the horizontal hohlraum axis) obtained with the 2 mg/cm$^3$ and 4 mg/cm$^3$ targets suggests that higher foam densities might lead to beam bending or Cross Beam Energy Transfer (CBET) \cite{CBET_multi_beam_review} from the inner to the outer beam cones.

In conclusion, foam-filled hohlraums show promising performance as an alternative solution to gas fills. Foams have been successfully demonstrated to provide the back pressure necessary to tamp the wall motion, whilst also producing similar amounts of backscattering to gas fills. This is encouraging as the amount of LPI can be further reduced by introducing dopants to the foam. Although offering slightly less back-pressure, low density foams (1 mg/cm$^3$) appear to be the best option as they generate less backscattering and best implosion symmetry. Finally, experimental results also show no reduction in performance in other crucial parameters when using foams, such as x-ray drive, and neutron yield, while keeping the production of hot-electrons relatively low -- making them an ideal candidate for a new high-gain hohlraum design on NIF.

\hfill\break
We would like to dedicate this work to the memory of our friend Oggie Jones whom has been a long-standing proponent of foams for the control of LPI instabilities in holraums.

\hfill\break
\textbf{Disclaimer:} This document was prepared as an account of work sponsored by an agency of the United States government. Neither the United States government nor Lawrence Livermore National Security, LLC, nor any of theur employees makes any warranty, expressed or implied, or assumes any legal liability or responsibility for the accuracy, completeness, or usefulness of any information, apparatus, product, or process disclosed, or represents that its use would not infringe privately owned rights. Reference herein to any specific commercial product, process, or service by trade name, trademark, manufacturer, or otherwise does not necessarily constitute or imply its endorsement, recommendation, or favouring the United States government or Lawrence Livermore National Security, LLC. The views and opinions of authors expressed herein do not necessarily state or reflect those of the United States government or Lawrence Livermore National Security, LLC, and shall not be used for advertising or product endorsement purposes.

This work was performed under the auspices of the U.S. Department of Energy by Lawrence Livermore National Laboratory under Contract DE-AC52-07NA27344 and supported by LDRD-21-ERD-041.

\bibliography{aipsamp}

\end{document}